\documentclass{appolb}
\usepackage{epsfig}
\usepackage{color}
\def\be{\begin{equation}}
\def\ee{\end{equation}}
\def\bea{\begin{eqnarray}}
\def\eea{\end{eqnarray}}
\definecolor{blueish}{rgb}{0.0,0.0,0.7}
\definecolor{greenish}{rgb}{0.0,0.7,0.0}
\definecolor{darkgreen}{rgb}{0.0,0.4,0.0}
\definecolor{turqoise}{rgb}{0.0,0.5,0.5}
\definecolor{gold}{rgb}{0.7,0.6,0.0}

\newcommand{\One}{1\!\!1}
\newcommand{\rse}{\mathcal{R}}
\newcommand{\bes}[1]{j^{#1}_{L_{#1}}}
\newcommand{\han}[1]{h^{(1)#1}_{L_{#1}}}

\newcommand{\tmat}[2]{T_{#1#2}^{(L_{#1},L_{#2})}}

\begin{document}
\title{Light and Not So Light Scalar Mesons
\thanks{Talk at Workshop \em ``Excited QCD'', \em Zakopane, Poland,
8--14 Feb.\ 2009}
}
\author{George Rupp$^*$, Susana Coito
\address{
Centro de F\'{\i}sica das Interac\c{c}\~{o}es Fundamentais,
Instituto Superior T\'{e}cnico, Technical University of
Lisbon, P-1049-001 Lisboa, Portugal} 
\and
Eef van Beveren
\address{
Centro de F\'{\i}sica Computacional, Departamento de F\'{\i}sica,
Universidade de Coimbra, P-3004-516 Coimbra, Portugal}
}
\maketitle
\begin{abstract}
A multichannel description of the light scalar mesons in the framework of
the Resonance-Spectrum Expansion is generalised by including vector-vector
and scalar-scalar channels, besides the usual pseudoscalar-pseudoscalar
channels. Experimental data for the isoscalar, isodoublet and isovector
cases are fitted up to energies well above 1~GeV. The resulting pole 
positions of the light and intermediate scalar mesons are compared to
the listed resonances. Possible further improvements are discussed.
\end{abstract}

\section{Introduction} 
The light scalar mesons represent nowadays one of the hottest topics in
hadronic physics. Despite the growing consensus on the existence of a
complete light scalar nonet, comprising the $f_0$(600) (alias $\sigma$),
$K_0^*$(800) (alias $\kappa$), $a_0$(980) and $f_0$(980), which are now all
included \cite{PDG08} in the PDG tables, their interpretation and possible
dynamical origin in the context of QCD-inspired methods and models remains
controversial. Moreover, their classification with respect to the intermediate
scalars $f_0$(1370), $K_0^*$(1430), $a_0$(1450) and $f_0$(1500) \cite{PDG08}
is also subject to continued debate. For a brief historical discussion of the
main theoretical and phenomenological approaches to the light scalars and the
corresponding references, see Refs.~\cite{0812.1527}.

In the present work, the successful multichannel description of the light
scalar mesons in Ref.~\cite{BBKR06} is further generalised by including,
besides the usual pseudoscalar-pseudoscalar (PP) channels, also all
vector-vector (VV) \cite{0812.1527} and scalar-scalar (SS) channels comprising
light mesons. This is crucial for an extension of the applicability of the
approach to energies well above 1~GeV, so as to make more reliable predictions
for the intermediate scalar resonances as well.  

\section{Resonance-Spectrum Expansion}

\begin{figure}[h]
\begin{tabular}{lr}
\epsfysize=45pt
\epsfbox{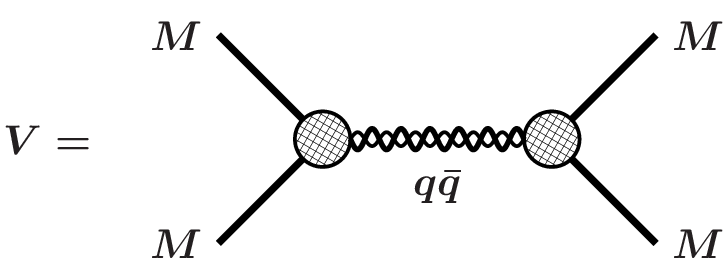}
&
\hspace*{10pt}
\epsfysize=45pt
\epsfbox{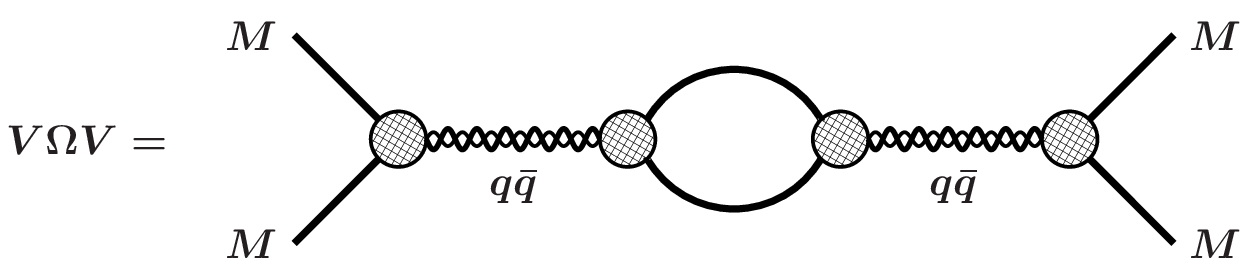} 
\end{tabular}
\caption{Born (left) and one-loop (right) term of the RSE effective
meson-meson interaction (see text).}
\label{bornoneloop}
\end{figure}

We shall study the scalar mesons in the framework of the
Resonance-Spectrum-Expansion (RSE) model \cite{BR06a}, in which mesons in
non-exotic channels scatter via an infinite set of intermediate $s$-channel
$q\bar{q}$ states, i.e., a kind of Regge propagators \cite{BR08}. The
confinement spectrum for these bare $q\bar{q}$ states can, in principle, be
chosen freely, but in all successful phenomenological applications so far we
have used a harmonic-oscillator (HO) spectrum with flavour-indepedent
frequency, as in Refs.~\cite{BRRD83} and \cite{ERMDRR86}. Because of the
separability of the effective meson-meson interaction, the RSE model can be
solved in closed form. The relevant Born and one-loop diagrams are depicted in
Fig.~\ref{bornoneloop},
from which it is obvious that one can straightforwardly sum up the complete
Born series. For the meson-meson--quark-antiquark vertex functions we take a
delta shell in coordinate space, which amounts to a spherical Bessel function
in momentum space. Such a transition potential represents the breaking of
the string between a quark and an antiquark at a certain distance $a$, with
overall coupling strength $\lambda$, in the context of the $^{3\!}P_0$ model.
The fully off-energy-shell $T$-matrix can then be solved as
\[
\tmat{i}{j}(p_i,p'_j;E)=-2a\lambda^2\,U_i\,\bes{i}(p_ia)\!\sum_{m=1}^{N}\!
\rse_{im}\left\{[\One-\Omega\,\mathcal{R}]^{-1}\right\}_{\!mj}
\bes{j}(p'_ja)\,U'_j,
\label{tfinal}
\]
with
\[
\mathcal{R}_{ij}(E)=\sum_{\alpha}\sum_{n=0}^{\infty}
\frac{g_i^{(\alpha,n)}g_j^{(\alpha,n)}}{E-E_n^{(\alpha)}}\;,\;\;\;
\Omega_{ij}=-2ia\lambda^2\mu_jk_j\,\bes{j}(k_ja)\,\han{j}(k_ja)\,\delta_{ij}\;, 
\]
and where $E_n^{(\alpha)}$ is the discrete energy of the $n$-th recurrence in
$q\bar{q}$ channel $\alpha$, $g_i^{(\alpha,n)}$ is the corresponding coupling
to the $i$-th meson-meson channel, $k_j$ and $\mu_j$ are the (relativistically
defined) on-shell momentum and reduced mass of meson-meson channel $j$,
respectively, $\bes{j}(k_ja)$ and $\han{j}(k_ja)$ are the $L_j$-th order
spherical Bessel function and spherical Hankel function of the first kind,
respectively, and $U_i=\sqrt{\mu_ip_i}$.

Spectroscopic applications of the RSE are manifold. In the one-channel
formalism, the $\kappa$ meson was once again predicted, before its experimental
confirmation, in Ref.~\cite{BR01}, 1st paper, after its much earlier prediction
in Ref.~\cite{ERMDRR86}. In the 2nd paper of Ref.~\cite{BR01}, the low mass of
the $D_{s0}^*$(2317) was shown to be due to its strong coupling to the $S$-wave
$DK$ threshold, an explanation that is now widely accepted. The 3rd paper of
Ref.~\cite{BR01} presented a similar solution to the whole pattern of masses
and widths of the charmed axial-vector mesons.

Multichannel versions of the RSE model have been employed to produce
a detailed fit of $S$-wave PP scattering and a complete light scalar nonet
\cite{BBKR06}, with very few parameters (also see below),
and to predict the $D_{sJ}$(2860) \cite{BR06b}, shortly
before its observation was publicly announced.

Finally, the RSE has recently been applied to production processes \cite{BR07}
as well, in the spectator approximation. Most notably, it was shown that the
RSE results in a \em complex \em \/relation between production and scattering
amplitudes (papers 1--3 in Ref.~\cite{BR07}). Successful applications include
the extraction of $\kappa$ and $\sigma$ signals from data on 3-body decay
processes (4th paper in Ref.~\cite{BR07}), the deduction of the string-breaking
radius $a$ from production processes at very different energy scales
(5th paper), and even the discovery of signals hinting at new vector charmonium
states in $e^+e^-\to\Lambda_c\bar{\Lambda}_c$ data (6th paper). 
\section{Light and Intermediate Scalar Mesons} 
\subsection{Published results for $S$-wave PP scattering}
In Ref.~\cite{BBKR06}, two of us
(E.v.B, G.R.) together with Bugg and Kleefeld applied the RSE to $S$-wave PP
scattering up to 1.2 GeV, coupling the channels
$\pi\pi$, $K\bar{K}$, $\eta\eta$, $\eta\eta'$, $\eta'\eta'$ for $I\!=\!0$,
$K\pi$, $K\eta$, $K\eta'$ for $I\!=\!1/2$, and $\eta\pi$, $K\bar{K}$,
$\eta'\pi$ for $I\!=\!1$. Moreover, in the isoscalar case both an $n\bar{n}$
and an $s\bar{s}$ channel were included, so as to allow dynamical mixing to
occur via the $K\bar{K}$ channel. The very few parameters, essentially only
the overall coupling $\lambda$ and the transition radius $a$, were fitted to
scattering data from various sources, for $I\!=\!0$ and for $I\!=\!1/2$, and to
the $a_0$(980) line shape, determined in a previous analysis, for $I\!=\!1$.
Moreover, the parameters $\lambda$ and $a$ varied less than $\pm10\%$ from
one case to another. Overall, a good description of the data was achieved
(see Ref.~\cite{BBKR06} for details). Poles for the light scalar mesons were
found at (all in MeV) \\[1.5mm] \indent
$\sigma\!: 530-i226\;, \;\;
\kappa\!: 745-i316\;, \;\;
f_0(980)\!: 1007-i38\;, \;\;
a_0(980)\!: 1021-i47 \;.$ \\[1.5mm]
No pole positions for the intermediate scalars were reported in
Ref.~\cite{BBKR06}, as the fits were only carried out to 1.1 GeV in the
isovector case, and to 1.2 GeV in the others. Nevertheless, corresponding
poles at higher energies were found, but these were of course quite
unreliable.

In the following, we shall present preliminary results for fits extended
to higher energies, and with more channels included.

\subsection{Isoscalar scalar resonances with PP, VV and SS channels included}
\begin{figure}[b]
\begin{tabular}{lr}
\epsfxsize=175pt
\hspace*{-20pt}
\epsfbox{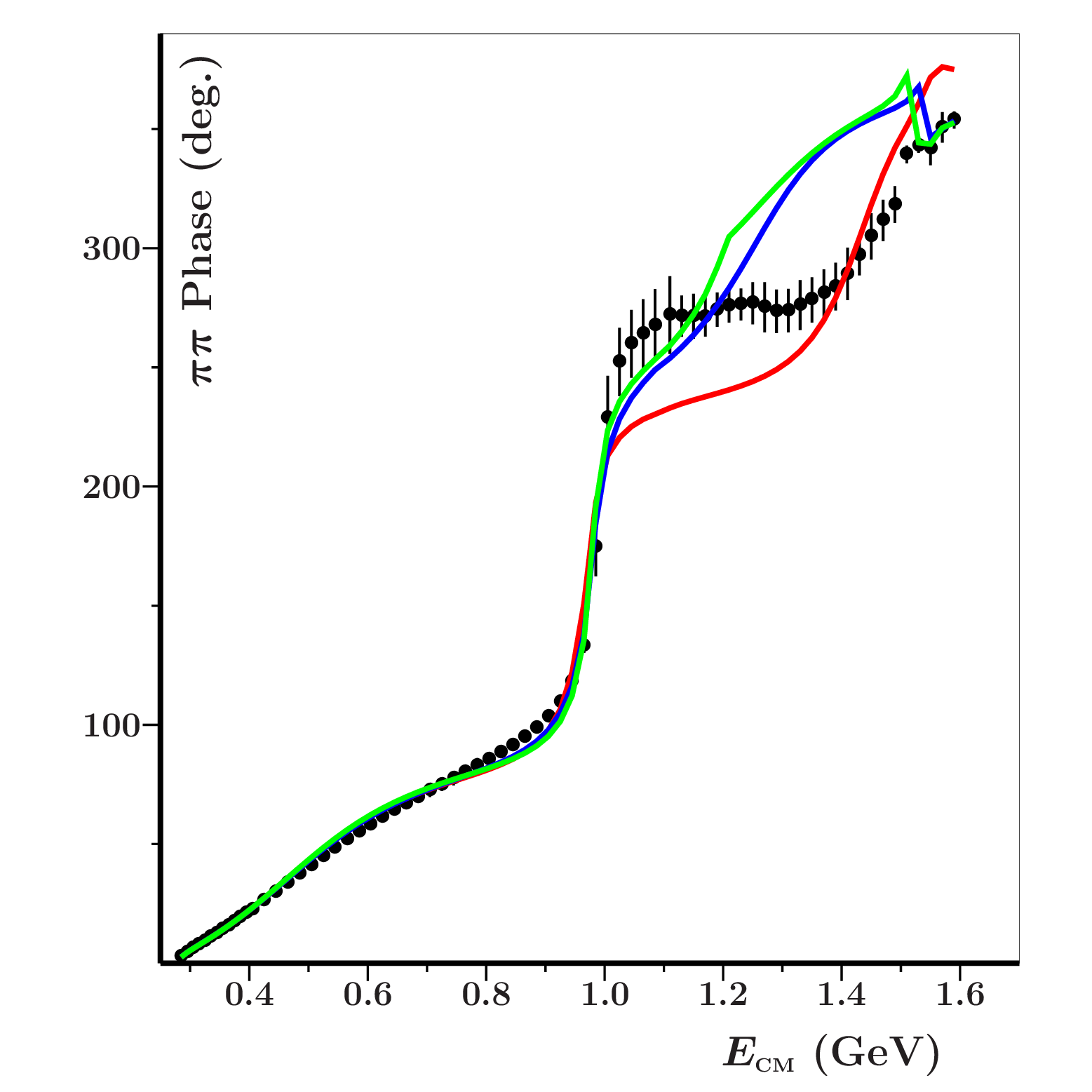}
&
\epsfxsize=175pt
\epsfbox{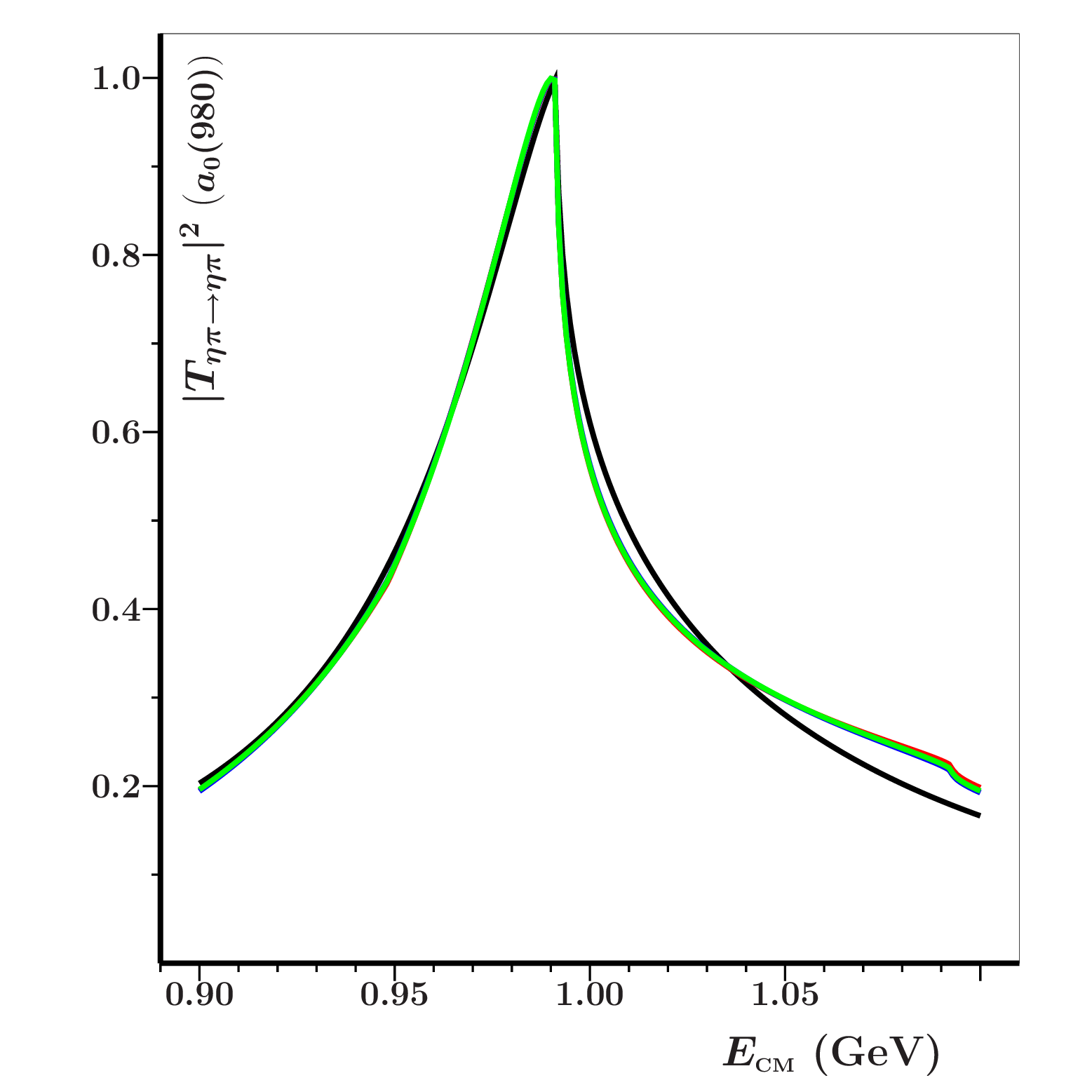}
\end{tabular}
\caption{Left: $S$-wave $I\!=0\!$ $\pi\pi$ phase shifts; red: PP channels
only; blue: also VV channels; green: also SS channels; data due to
Ref.~\cite{BS08}. 
Right: $a_0(980)$ line shape; coloured curves: as in left-hand plot; black
curve from data analysis by Bugg \cite{BS08}.}
\label{sigmaf0a0}
\end{figure}
For $I\!=\!0$, the VV channels that couple to $n\bar{n}$ and/or $s\bar{s}$
are $\rho\rho$, $\omega\omega$, $K^*\bar{K}^*$ and $\phi\phi$, for both
$L\!=\!0$ and $L\!=\!2$, while the SS channels are $\sigma\sigma$,
$f_0(980)f_0(980)$, $\kappa\kappa$ and $a_0(980)a_0(980)$, with $L\!=\!0$ only.
We fit the parameters $\lambda$ and $a$ to sets of $S$-wave $\pi\pi$ phase
shifts compiled by Bugg and Surov\-tsev \cite{BS08}, which yield a somewhat
larger scattering length than in Ref.~\cite{BBKR06}, viz.\ $0.21m_\pi^{-1}$.
The results of the fits are shown in Fig.~\ref{sigmaf0a0}, left-hand plot,
together with the curve from Ref.~\cite{BBKR06}, where only PP channels were
included and
somewhat lower data were used just above the $\pi\pi$ threshold. All three fits
are good up to 1~GeV, with only small differences among them. Thereabove, the
VV and SS channels clearly produce very substantial effects, though
overshooting between 1.2 and 1.6~GeV. Possible improvements are discussed
below. With all channels included, we find the first four isoscalar poles at
(all in MeV) \\[1.5mm]
\indent
$\sigma\!: 464-i217\;, \;\;\;
f_0(980)\!: 987-i29\;, \;\;\;
f_0(1370)\!: 1334-i185\;$, \\[1mm] \indent
$f_0(1500)\!: 1530-i14 \;.$ \\[1mm]
Moreover, there is an extra broad state at $(1519-i219)$ MeV, probably of a
dynamical origin. The present predictions for the $f_0$(1370) and the
$f_0$(1500) are clear improvements with respect to the case with PP and VV
channels only (see Ref.~\cite{0812.1527}, 1st paper). In particular, the extra
pole found here might help to explain the experimental difficulties 
with the $f_0$(1370) and $f_0$(1500).
\subsection{\boldmath{$a_0(980)$} and \boldmath{$a_0(1450)$}}
In the isotriplet case, we fit $\lambda$ and $a$, as well as
the pseudoscalar mixing angle, to the $a_0$(980) line shape, just as in
Ref.~\cite{BBKR06}, but now with the VV ($\rho K^*$, $\omega K^*$, $\phi K^*$)
and SS ($a_0(980)\sigma$, $a_0(980)f_0(980)$, $\kappa\kappa$) channels added.
Thus, the quality of the fit is slightly improved, though the differences with
the PP and PP+VV cases are hardly visible in Fig.~\ref{sigmaf0a0}, right-hand
plot. The poles we find are $(1023-i47)$~MeV (second sheet) for the $a_0$(980)
and $(1420-i185)$~MeV for the $a_0$(1450), which are very reasonable values 
\cite{PDG08}.
\subsection{\boldmath{$K_0^*(800)$} and \boldmath{$K_0^*(1430)$}}
Adding the vector ($\rho K^*$, $\omega K^*$, $\phi K^*$) and scalar
($\sigma\kappa$, $f_0(980)\kappa$, $a_0(980)\kappa$) channels in the
isodoublet sector does not allow a stable fit to be obtained. Moreover, the
LASS data are known to violate unitarity above 1.3 GeV. So we just fit up to
1.3~GeV, with only the PP channels, getting parameters very close to the
isoscalar case, including a reasonable pseudoscalar mixing angle. See further
the conclusions for possible remedies. From the present PP fit we find the
pole postions $(722-i266)$~MeV for the $\kappa$ and $(1400-i96)$~MeV for the
$K_0^*$(1430), which are again reasonable values \cite{PDG08}.
\section{Conclusions and outlook}
The preliminary results in this study indicate that a good description of both
the light and the intermediate scalar mesons is feasible in the RSE, by
taking into account additional sets of coupled channels that should become
relevant at higher energies. However, some problems persist, like the too slow
rise and subsequent overshooting of the $I\!=\!0$ $\pi\pi$ phase shift above
1~GeV, and the mentioned fitting problems in the isodoublet case. A possible
cause of these difficulties is the assumed sharpness of several thresholds
involving broad resonances in their turn, such as $\sigma\sigma$, $\rho\rho$,
$\sigma\kappa$, etc., which may result in too drastic effects at the opening
of these channels. We are now studying ways to account for final-state
resonances having non-zero widths, without destroying unitarity. It might also
turn out to be necessary to consider more general transition potentials.

\section*{Acknowledgements}
We are grateful to the organisers for a most stimulating workshop. One of us
(G.R.) thanks J.~R.~Pel\'{a}ez for very useful information on $\pi\pi$
scattering. This work received partial financial support from the
{\it Funda\c{c}\~{a}o para a Ci\^{e}n\-cia e a Tecnologia}
\/of the {\it Minist\'{e}rio da Ci\^{e}ncia, Tecnologia e Ensino Superior}
\/of Portugal, under contract CERN/FP/83502/2008.

\end{document}